# Complex air-structured optical fibre drawn from a 3D-printed preform

Kevin Cook[1], John Canning[1*], Sergio Leon-Saval[2], Zane Reid[1,+], MD Arafat Hossain[1], Jade-Edouard Comatti[1], Yanhua Luo[3] and Gang-Ding Peng[3]

[1]*interdisciplinary Photonics Laboratories (iPL), School of Chemistry, The University of Sydney, NSW 2006, Australia;*
[2]*School of Physics, The University of Sydney, NSW 2006, Australia;*
[3] *National Fibre Facility, School of Electrical Engineering and Telecommunications, University of New South Wales, Sydney, NSW 2052, Australia;*
[+]*Currently with Precision 3D Printing (P3P), Unit 2, 28 Anvil Road, Silverdale, Auckland, New Zealand.*
*Corresponding author: john.canning@sydney.edu.au*

**A structured optical fibre is drawn from a 3D-printed structured preform. Preforms containing a single ring of holes around the core are fabricated using filament made from a modified butadiene polymer. More broadly, 3D printers capable of processing soft glasses, silica and other materials are likely to come on line in the not-so-distant future. 3D printing of optical preforms signals a new milestone in optical fibre manufacture.**

In recent years, there has been an explosion of interest in 3D printing technologies and there is now a vast range of 3D printing methods that are finding new applications every day in a whole host of areas from science and engineering to medicine and the arts [1]. They are set to revolutionise manufacturing. Fused deposition modelling (FDM) is one of the most commonly used techniques, being the first and simplest demonstration of 3D printing in the late 1980s [2]. It is especially suited for instrument prototype casings such as those required for novel smartphone spectrometers [3]. In FDM, a polymer is fed through a heated nozzle which melts the polymer - essentially a fancy thermal glue gun using a finite extrusion nozzle with programmable *xyz* positioning capability. Optical polymerisation methods using both light emitting diodes (LEDs) and lasers offer higher resolution and are increasingly popular; however, they do not have the same material range given the specific absorption and monomer polymerization requirements. Selective laser sintering (SLS) is another popular technique where a high power laser is utilised to fuse not only plastic, but glass, metal or ceramic particles or powders into 3D objects [4]. A 3D profile is achieved by scanning the laser on the horizontal plane layer-by-layer, lowering the printing bed between each layer. All these methods have their particular rate limiting steps determined by the consolidation process details and the need for full *xyz* scanning. More recently, holographic imaging using projectors promises significantly accelerated production of the 3D printed object by removing the requirement for *xy* scanning [5]. FDM, SLS and other related variants are experiencing tremendous growth world-wide, particularly for rapid prototyping and manufacturing. 3D printing is also being applied to cellular assembly processing of biomedical interest [6]. 3D printing technologies are reaching most fields, including more recently photonics. For example, a company called Luxexcel is already producing high transparency Fresnel lenses by printing with polymethylmethacrylate (PMMA) filament [7]. Most interesting, is recent work demonstrating direct-print short, solid plastic fibre [8] and "light pipes" [9] using transparent polymers, the first 3D printed waveguides.

Here, we explore harnessing 3D printing for optical fibre fabrication. We see it as revolutionising the manufacture of all optical fibre fabrication, including glass optical fibres as 3D glass printing comes on line [10,11]. Whilst there is debate as to the merits of the technology for conventional step-index optical fibres and how dopant distributions might be 3D printed, one breed of optical fibre that could particularly stand to benefit immediately are so-called structured (both microstructured and nanostructured) optical fibres. The fabrication of these fibres is limited to methods that are often suitable for specific materials. The stack-and-draw technique, where capillaries are manually assembled into a hexagonal structure prior to drawing [12], is the most popular method but is obviously limited to hexagonally packed periodic structures, giving rise to so-called photonic crystal fibres. More complex structures such as Fresnel [13,14] and related fractal fibres [15] cannot be easily constructed by assembling capillaries and have required other methods including drilling [12], extrusion [16,17] and injection molding [17]. As can be imagined, these methods are currently all time-consuming and 3D printing offers an immediately competitive advantage, and certainly for polymer fibres where 3D printing manufacturing is most mature.

In this paper, an alternative approach to making structured fibres is explored by utilising a 3D printer to design and print a structured preform that is then drawn to fibre. As proof of principle, the more mature FDM printing method is chosen to print a preform using a transparent thermosetting polymer that is subsequently drawn to fibre. An arbitrary fibre geometry consisting of a solid core surrounded by 6 air holes is chosen to assess the 3D-printed preform approach. The material used is a commercially available 3D printing filament consisting of a propriety polystyrene mixture containing styrene-butadiene-copolymer and polystyrene, labelled here as SBP. SBP can offer several significant advantages when compared to other polymer filaments: it is transparent, it has a high degree of flexibility and it does not suffer from discoloration under stress or when bent. A low-cost, commercially-available thermosetting 3D printer is used to print the fibre preform with this filament. The preform can be subsequently drawn to fibre with relative ease, without the need for pre-annealing or hole-pressurisation, maintaining its cross-sectional features after being drawn to fibre form. We also show that preform transparency can be improved by annealing which removes the scattering component arising from the printed interfacial layers and bubbles trapped during printing.

The aim of this paper is to demonstrate the first 3D printed structured optical fibres, choosing to focus on existing polymer based 3D printing as proof of concept. As new materials such as soft glasses and silica glasses become accessible to 3D printing, the revolution will eventually encompass all fibre technologies. In the meantime, structured polymer fibres have significant potential in their own right within a range of application areas. For example, chirped graded Fresnel fibre designs could find applications in LANs where step-index fibres are traditionally used [18]. Microstructured polymer optical fibre also shows strong potential in biomedical applications due to polymer compatibility with organic materials. Basic structured fibres with only a small number of holes offers cheap, biocompatible attenuation sensors for applications such as orthodontic pain monitoring within the human organism [19].

The air-structured preform was designed on Inventor computer aided design (CAD) software. Fig. 1 shows a depiction of the preform shape and a photo of the 3D printed preform end. The length of the preform was chosen to be $L$ = 10 cm, although given the volume workspace of the 3D printer used (a Deltasine "Redback-2"), lengths of up to $L$ = 39.5 cm can be achieved. The preform diameter of $\phi$ = 1.6 cm was chosen in order to be compatible with the polymer fibre draw tower and furnace. The solid core of the preform is surrounded by 6 air holes each with diameter $\phi_{hole}$ = 0.2 cm and evenly separated along the periphery. A low-cost, commercially-available 3D thermal printer was used to print the preform. The Redback-2 is a delta design model which is ideally-suited for printing circular structures compared to traditional *xyz* printers. The SBP polymer filament has a higher degree of transparency (specified > 90%, loss α < 1.2 dB) when compared to the more commonly used acrylonitrile butadiene styrene (ABS) polymer. The extruder and bed temperatures of the 3D printer were set at $T_{ex}$ = 210 °C and $T_{bed}$ = 95 °C respectively. The preform used in the following work was printed laterally with the preform written on it's side; however, it is also possible to print vertically. Resolution of the printing on the *xy*-axes was set to Δ*x* = Δ*y* = 0.4 mm and Δ*z* = 0.1 mm on the *z*-axis. For the case reported here the printing time was *t* = 6 hours (all automated, so no further steps need to be taken by the fabricator).

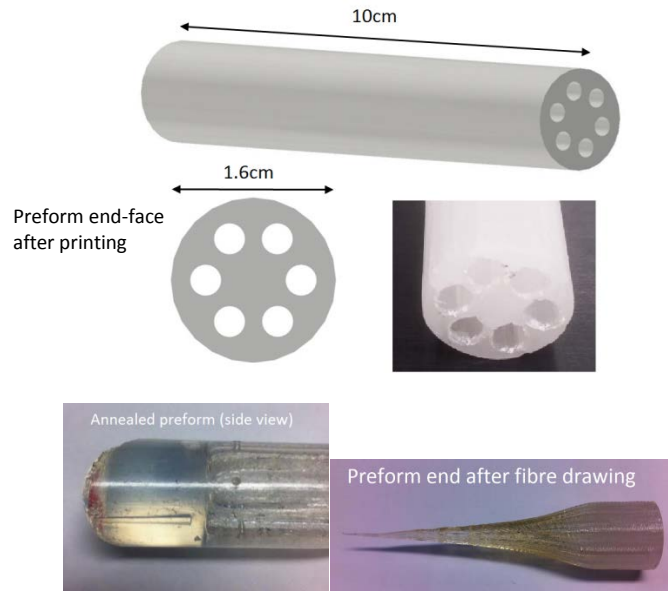

Fig. 1. Illustration of SBP fibre preform design and photos of annealed preform and preform after fibre draw.

Much of the observed opaqueness of the printed preform arises from scattering at the interface layer edges created by printing. However, it was found that drawing into optical fibre removed much of this and improved transparency such that post-annealing is not necessary. Annealing itself can remove significant scattering and dramatically improve transparency through the preform, making it easier to quantify attenuation and absorption spectroscopically. To demonstrate this, a second preform was printed, the only difference in design being that each end of the preform had a 2 cm long solid polymer end-cap in order help the preform maintain its circular shape during annealing. The preform was held in a computer-controlled oven under a vacuum ($P$ = 99 kPa, $T$ = 128 $^0$C, $t$ = 96 hrs). This was followed immediately by a secondary annealing step ($T$ = 148 $^0$C, $t$ = 48 hrs) and a third ($T$ = 168 $^0$C, $t$ = 96 hrs). After annealing, a 4 mm-thick slice was cut from the preform and polished using optical grade polishing paper. The transmission spectra were recorded in a spectrophotometer ($\lambda$ = 200 to 900 nm). Another identical sample was taken from a third, non-annealed preform and similarly polished for comparison. The transmission spectra in Fig. 2. (a) show a significant increase in transmission $Tr \sim$ 2 % (~43.9 dB/cm) to $Tr \sim$ 43 % (~9.5 dB/cm) at $\lambda$ = 900 nm after annealing. The results show increasing transmission in the near infrared (NIR). A Fourier Transform Infrared (FTIR) spectrometer was used to investigate transmission properties further into the NIR, Near IR-FTIR is a useful technique for the rapid and accurate identification of polymer materials [20]. Fig. 2. (b) shows the spectra for the annealed and non-annealed samples. The characteristic C-H overtone bands of polystyrene are clearly-discernable at 1.140 nm and 1.209 nm, as is significant O-H absorption in the region of 1.415 nm. Most notable, however, is the increase in transmission over the spectral window of 1500 – 1600 nm; this is an important transmission band for telecommunications. Above this region the transmission is observed to fall-off rapidly, a common feature exhibited by other polymers such as acrylic and polycarbonate-based materials [21].

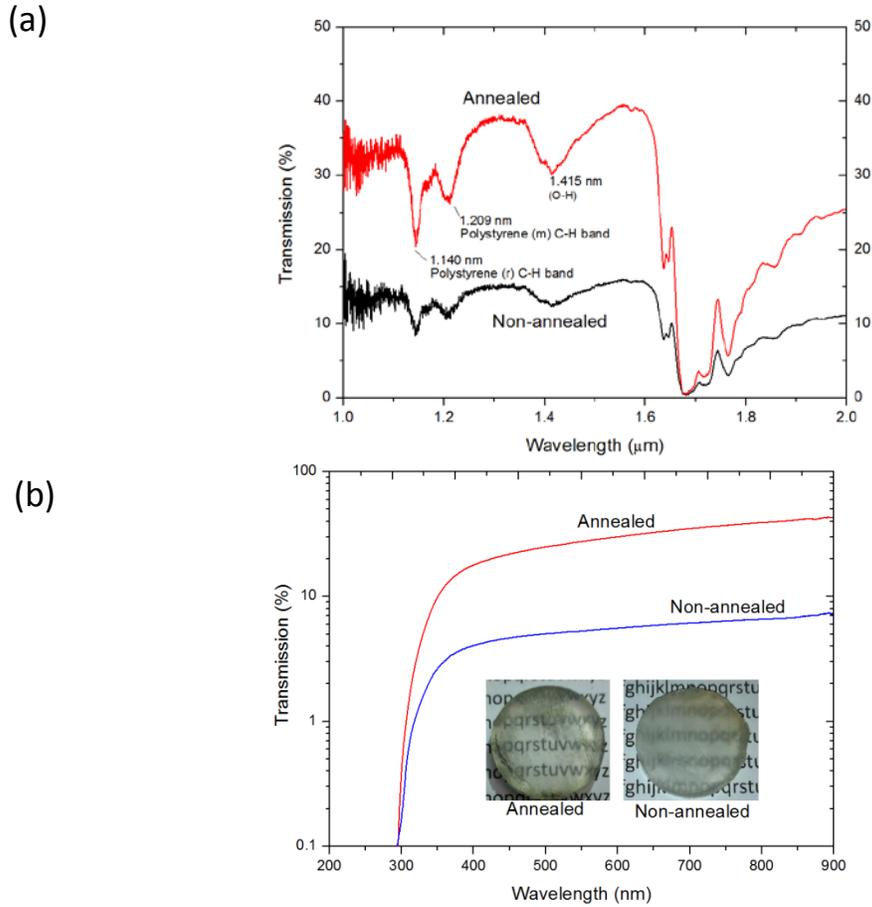

Fig. 2. Transmission (*Tr*) spectra and photos of the annealed and non-annealed preform: (a) UV-Visible spectra taken with a spectrophotometer; (b) IR spectra taken using a FTIR spectrometer indicating 2 polystyrene overtones for the aryl (r) and methelnic (m) C-H bands as well as a characteristic O-H band.

The air-structured preform was drawn into fibre using a draw tower with a low-temperature radiative furnace. The preform is passed through the furnace at a feed speed $v$ = 1 mm/min. No gases were required to pressurise the holes and the temperature of the furnace was stabilized at $T$ = 160 $^0$C. The preform was pulled out of the furnace into fibre at a rate $v$ = 35 mm/min. Once the process is optmised, it was possible to spool a fibre length of up to $L \sim 10$ m. The preform experienced a small degree of bulging (so-called "mushrooming") inside the furnace during the first few minutes of drawing. This deformation led to the preform becoming slightly elliptical in shape. The likely causes involve incomplete polymerisation during printing as well as trapped air and moisture.

Optical microscope images of the final fibre cross section are shown Fig. 3. (a) and (b). A length of fibre $L$ = 15 cm is cleaved and then illuminated from one end by a broadband white light source. The other end of the fibre is then imaged using red and green bandpass filters. A number of air bubbles are clearly discernable in these images, but improvements on printer settings such as resolution and printing speeds, as well as additional preform pre-annealing, can help to eliminate or minimise these defects. The outer diameter of the fibre is measured to be $\emptyset_{OD1} \sim 712$ μm and $\emptyset_{OD2} \sim 605$ μm for the major and minor axes respectively, whilst the core diameter is $\emptyset_{c1} \sim 221$ μm and $\emptyset_{c2} \sim 148$ μm for the major and minor axes respectively. A photo of white light output of the fibre projected onto a screen $\sim$ 1cm from the fibre end-face is shown in Fig. 3. (c). In order to assess the optical guidance properties of the fibre, the output of a HeNe laser ($\lambda$ = 632.8 nm), a Nd:YAG laser ($\lambda$ = 1064 nm) and the line of a tunable semiconductor laser ($\lambda$= 1550 nm) were each, in turn, coupled into the fibre and the output power measured. The fibre was fixed in place so as to make a 90° turn with a radius of curvature $r \sim 6$ cm. This simple experiment reveals upper estimates of the propagation loss, $\alpha$, at each wavelength: $\alpha \sim 1.5$ dB/cm @ 632 nm, $\alpha \sim 0.75$ dB/cm @ 1064 nm, and $\alpha \sim 1.51$ dB/cm @ 1550 nm. These values compare favorably to structured optical fibres produced by capillary stacking and drilling, particularly given the large interstitial spacing between holes [22] suggesting that confinement loss is curtailed by the outer diameter of the fibre where a large number of higher order modes are supported. Ideally, confinement losses estimated for a single ring of holes in a lossless material are on the order of $\alpha \sim 0.2$ dB/cm for the fundamental mode and $\alpha \sim 2 \times 10^4$ dB/cm for the first higher order mode [23], so significant scope for improvement remains. Noticeably these values are a significant improvement over the preform attenuations even after annealing, closer to the ideal losses expected based on material properties.

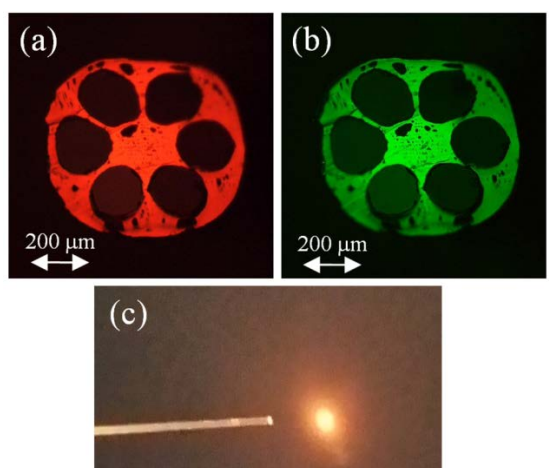

Fig. 3. Transmission of light through the drawn optical fibre: (a) 630 nm and (b) 515 nm guided light imaged at the fibre end face; (c) white light output projected onto a screen.

In conclusion the first design and fabrication of an optical fibre using a preform made with 3D printing technology has been demonstrated. The approach is simple and harnesses the power of a 3D printer to print shapes of arbitrary geometries, avoiding the problems often associated with traditional preform fabrication. Although the printing time was several hours and the printing resolution low using a low-cost thermal based printer, it was automated and there were no additional steps required by an operator, achieving a huge saving in person-hours. More expensive printers, other types of printers such as SLED, sintering and next generation holographic slicer printers that allow real time fabrication of the preform will accelerate production to unprecedented levels. Other materials such as silica will soon come on line; we have shown 3D printing will revolutionise fibre preform manufacture and the potential exists to eventually enable 3D printing of fibre itself. With minimal preparation time required beforehand using user-friendly CAD software (Inventor), a fabricator could easily design and start printing an air-structured preform within 30 minutes or so depending on the complexity of the design. The ability to fabricate *custom* 3D printing filaments made possible by extruder technology is also another exciting possibility and gives scope to easily introduce dopants into fibres; the use of multi extruder tips will allow multiple materials to be combined within the one preform, including multiple cores and much more. Of more immediate benefit to plastic research for biosensing, terahertz, metamaterials and so on, we have also shown that the modified butadiene plastic is suitable for polymer optical fibre work.

**Funding.** This project was primarily funded by the Australian Research Council (ARC) through Grants FT110100116 and DP140100975. The National Fibre Facility is funded by ARC grants LE0883038 and LE100100098.

**Acknowledgments**. Z. Reid acknowledges an *i*PL Summer scholarship; Md. Hossain acknowledges an Australian Postgraduate Award. This work was performed in part at the Optofab node of the Australian National Fabrication Facility (ANFF) using Commonwealth and NSW State Government funding.